\documentclass[preprint,5p]{elsarticle}

\usepackage{pifont,geometry,fleqn}
\usepackage{amsfonts,amsmath,amssymb}
\usepackage{graphicx,float,color}
\usepackage{hyperref}
\hypersetup{colorlinks = true, linkcolor = blue, urlcolor  = blue, citecolor = blue, anchorcolor = blue}
\usepackage{multirow}
\usepackage{braket}
\usepackage{epic,eepic,epsfig,latexsym}

\usepackage{breakurl}  
\usepackage[caption=false]{subfig}
\usepackage{bm}
\biboptions{sort&compress}
\usepackage{lipsum}
\makeatletter
\def\ps@pprintTitle{%
 \let\@oddhead\@empty
 \let\@evenhead\@empty
 \def\@oddfoot{}%
 \let\@evenfoot\@oddfoot}
\makeatother

\usepackage{etoolbox}
\patchcmd{\abstract}{Abstract}{ \hspace{7.75cm} Abstract}{}{}
 
\begin{document}
\newgeometry{left=1.8cm,right=1.5cm,bottom=2.5cm,top=2.8cm} 
\title{\textbf{ A scale at 10 MeV, gravitational topological vacuum, \\and large extra dimensions}}

\author{Ufuk Aydemir \corref{cor1}} 
\cortext[cor1]{\textit{Email address:} uaydemir@hust.edu.cn}

\address{School of Physics,
Huazhong University of Science and Technology, Wuhan, Hubei 430074, P. R. China**\corref{cor2},}
\cortext[cor2]{Current affiliation}
\address{Department of Physics and Astronomy,
Uppsala University, Box 516, SE-75120 Uppsala, Sweden \vspace{-0.4cm}}

\date{\today}
\begin{abstract}
We discuss a possible scale of gravitational origin at around $10$ MeV, or $10^{-12}$ cm, which arises in the MacDowell-Mansouri formalism of gravity due to the topological Gauss-Bonnet term in the action, as pointed out by Bjorken several years ago.~A length scale of the same size emerges also in the Kodama solution in gravity, which is known to be closely related to the MacDowell-Mansouri formulation.~We particularly draw attention to the intriguing incident that existence of six compact extra dimensions originated from TeV-scale quantum gravity as well points to a length scale of $10^{-12}$ cm, as the compactification scale.~The presence of six such extra dimensions is also in remarkable consistency with the MacDowell-Mansouri formalism; it provides a possible explanation for the factor of $\sim10^{120}$ multiplying the Gauss-Bonnet term in the action. We also comment on the relevant implications of such a scale regarding the thermal history of the universe motivated by the fact that it is considerably close to $1-2$ MeV below which the weak interactions freeze out, leading to Big Bang Nucleosynthesis. 
\end{abstract}

\begin{keyword}   
MacDowell-Mansouri formalism, Bjorken-Zeldovich scale, Gauss-Bonnet term, Einstein-Cartan formalism, Big Bang Nucleosynthesis, cosmological constant, Kodama wavefunctions\end{keyword}

\maketitle

\section{Introduction\label{sec:intro}}

 Bjorken points out in Ref.~\cite{ANDP:ANDP201300724} that the MacDowell-Mansouri (MM) formulation of gravity~\cite{MacDowell:1977jt} naturally reveals an induced scale of  $\sim 10$ MeV, or $\sim10^{-12}$ cm, which he names after Zeldovich, inspired by Zeldovich's seminal papers~\cite{Zeldovich:1967gd,Zeldovich:1968ehl}.~The MM formulation unifies the tetrad and spin connection of the first order Einstein-Cartan formalism, which take values in $SO(3,1)$, into a grand connection that lives in $SO(4,1)$ (or $SO(3,2)$ for a negative cosmological constant). The resulting action, through breaking the $SO(4,1)$ symmetry down to the   $SO(3,1)$, yields the usual Einstein-Hilbert term, a cosmological constant, and the Gauss-Bonnet (GB) term, which is topological in four dimensions~\cite{Freidel:2005ak,Wise:2006sm,Wise:2009fu}.
\raggedbottom

Intriguingly, a length scale of $10^{-12}$ cm, as noted in Ref.~\cite{Randono:2008wt}, is also encountered in the context of so-called the Kodama wavefunction in gravity~\cite{Kodama:1988yf,Smolin:2002sz,Witten:2003mb, Randono:2006rt,Randono:2006ru}, analogous to the Chern-Simons solution in Yang-Mills theory in four dimensions, which is also an important element in Loop Quantum Gravity~\cite{Rovelli:2010bf,Ashtekar:2017iip}. Actually, there is known to be a connection between the inner product of Kodama states and the MM formalism; Ref.~\cite{Randono:2006ru} points out that the topological terms arisen in the (extended) MM action and the inner product are the same. 

Bjorken, additionally, suggests six extra spatial dimensions, assumed to be compactified on this induced scale of $10^{-12}$ cm, simply to account for the large factor multiplying the MM action~\cite{ANDP:ANDP201300724}; $\sim10^{120}$, which, quite remarkably, also happens to be the infamous number often encountered in the cosmological constant problem~\cite{Weinberg:1988cp,Carroll:2000fy,Polchinski:2006gy,Padilla:2015aaa}.   

In this paper, we emphasize that the TeV-scale quantum gravity picture with large extra dimensions (LED) \cite{ArkaniHamed:1998rs,Antoniadis:1998ig,ArkaniHamed:1998nn,ArkaniHamed:2000eg,Antoniadis:1997zg,Accomando:1999sj,Antoniadis:1999bq} (known as the ADD model) naturally reveals a scale of $\sim10^{-12}$ cm as the compactification scale, provided that the number of extra spatial dimensions is set to six, with no need for an ad-hoc assumption of the corresponding length scale. In order to be consistent with the known physics up to the TeV-scale, we adopt the well-known approach that only the graviton is allowed to propagate throughout the bulk experiencing the extra dimensions, while the Standard Model (SM) fields are localized to the usual 4 dimensions. This, in this scenario, would introduce a deviation in the gravitational interactions on scales smaller than $10^{-12}$ cm;  the gravitational interaction has so far been tested down to the scale of $0.01$ cm~\cite{Kapner:2006si}.
 
Moreover, we notice a combination of the ``Bjorken-Zeldovich (BZ) scale" and the TeV scale, $M_{BZ}^3/M_{EW}^2\sim 10^{-3}$ eV, which is in the order of the observed vacuum energy density in the present universe and in the ballpark of the anticipated neutrino masses~\cite{Ade:2015xua,Olive:2016xmw}. Although it is most likely a coincidence, we present several toy models which illustrate its possible role as some sort of a see-saw-type suppression in obtaining the neutrino 
\restoregeometry  
\hspace{-0.64cm} 
mass and the cosmological constant. 

We also comment on possible other implications in cosmology.~This scale is considerably close to $1-2$ MeV below which the weak interactions freeze out, leading to Big Bang Nucleosynthesis (BBN). Premised on our current understanding of BBN, it is in general supposed that any deviation from the known radiation density around the decoupling temperature would change the time scale associated with BBN, and it is thus tightly constrained from the observations on the primordial abundances of light elements~\cite{Carroll:2004st,Weinberg:2008zzc,Langacker:2010zza}.\footnote{Recently, Atomki group in Hungary has reported an anomaly in the $^{8}$Be nuclear decay by internal $e^+ e^-$ formation at an invariant mass $m_{\overline{e}e}\cong17$ MeV, with a statistical significance of 6.8$\sigma$~\cite{Krasznahorkay:2015iga}. See also Refs.~\cite{deBoer:1996qdk,deBoer:1997mr,deBoer:2001sjo,deBoer:2005kf,Krasznahorkay:2005jy,Krasznahorkay:2006zf,deBoer:2010qb,Wojtsekhowski:2012zq,Gulyas:2015mia} for the previous studies relevant to this observation. The observation has ignited interest in the high energy physics community to suggest explanations some of which consider a hidden sector at around this energy scale whose effects have so far remained unnoticed~\cite{Feng:2016jff,Feng:2016ysn,Gninenko:2016kpg,Gu:2016ege,Jia:2016uxs,Kitahara:2016zyb,Ellwanger:2016wfe,Chen:2016tdz,Neves:2016nek,Kahn:2016vjr,Fayet:2016nyc,Neves:2016ugb,Kozaczuk:2016nma,Chiang:2016cyf,Krasnikov:2017dmg,Araki:2017wyg,Benavides:2016utf,Neves:2017rcn,DelleRose:2017xil}.}
 
\section{The MacDowell-Mansouri formalism and the \\Bjorken - Zeldovich scale}

Bjorken, in Ref.~\cite{ANDP:ANDP201300724}, discusses how a scale of $\sim 10$ MeV is induced in the MacDowell-Mansouri formalism through the Gauss-Bonnet (GB) topological term arisen naturally in the formalism in addition to the usual Einstein-Hilbert action and a cosmological constant term. 

The $SO(3,1)$ MM action, obtained through breaking the $SO(4,1)$ symmetry, is given as~\cite{MacDowell:1977jt,ANDP:ANDP201300724,Freidel:2005ak,Wise:2006sm,Wise:2009fu} 
\begin{flalign}\label{action1}
\mathcal{S}_{MM}=\frac{M_{Pl}^2}{64\pi H_0^2}\int d^4 x \sqrt{-g}\ \frac{1}{4} \; F^{ab}_{\mu\nu} F^{cd}_{\lambda\sigma}\;\epsilon_{abcd}\;\epsilon^{\mu\nu\lambda\sigma}\;,
\end{flalign}
where the $\epsilon$ symbols denote Levi-Civita tensors, $F^{ab}_{\mu\nu}= R_{\mu\nu}^{ab}-H_0^2 \left(e_{\mu}^a e_{\nu}^b -e_{\nu}^a e_{\mu}^b \right)$, $R_{\mu\nu}^{ab}=R_{\mu\nu}^{\rho\sigma} e_{\rho}^a e_{\sigma}^b$ is the Riemann tensor, $e_{\mu}^a$ is the tetrad (vielbein), and $a,\mu=0,1,2,3$ are the indices of the internal $SO(3,1)$ space and the four dimensional space-time, respectively. $H_0$ is the Hubble constant. 

Note that $F^{ab}_{\mu\nu}$ is the $SO(3,1)$ projection of the curvature $F^{AB}_{\mu\nu}$, constructed from the generalized connection $A^{AB}_{\mu}$ ($A=0,1,2,3,4$) that lives in a local $SO(4,1)$.~$A^{AB}_{\mu}$ takes the following form.$\;A^{4a}_{\mu}\equiv H_0 e^{a}_{\mu}$ and $A^{ab}_{\mu}\equiv w^{ab}_{\mu}$, where $w$ is the spin connection which lives in the $SO(3,1)$ group.

The action in Eq.~(\ref{action1}) yields
\begin{eqnarray}\label{action2}
\mathcal{S}_{MM}&=&\frac{M_{Pl}^2}{8\pi}\int d^4 x \sqrt{-g}\\
&&\Bigg(\frac{1}{32H_0^2} R^{\alpha\beta}_{\mu\nu}R^{\gamma\delta}_{\lambda\sigma} \epsilon_{\alpha\beta\gamma\delta} \;\epsilon^{\mu\nu\lambda\sigma}
+\frac{1}{2} R- \Lambda\Bigg)\;,\nonumber
\end{eqnarray}
where the cosmological constant $\Lambda = 3 H_0^2$ as it ought to be, and the first two terms are the GB and the Einstein-Hilbert terms, respectively. The GB term can be written in the more familiar form as
\begin{flalign}
\frac{1}{4} R^{\alpha\beta}_{\mu\nu}R^{\gamma\delta}_{\lambda\sigma} \epsilon_{\alpha\beta\gamma\delta} \;\epsilon^{\mu\nu\lambda\sigma}
=- \left( R^{\alpha\beta}_{\mu\nu}R^{\mu\nu}_{\alpha\beta}-4R^{\alpha\beta} R_{\alpha\beta}+R^2\right).
\end{flalign}

Notice the factor $\sim 10^{120}$ in front of the GB term in Eq.~(\ref{action2}), also multiplying the total MM action in Eq.~(\ref{action1}), which happens to be the infamous number in the cosmological constant problem ($\frac{M_{Pl}^4}{64\pi^2 \rho_{\Lambda}}\sim 10^{120}$). The possible role of this factor of the MM action in the resolution of the cosmological constant problem has not been demonstrated yet, to the best of our knowledge. 


In the Friedmann-Robertson-Walker (FRW) background, where
\begin{equation}
ds^2=-dt^2+a^2(t) dx_i dx_i\;,
\end{equation}
 the GB term in Eq.~(\ref{action2}) becomes
\begin{equation}\label{GBFRW}
\mathcal{S}_{GB}=-\frac{M_{Pl}^2 V(0)}{8\pi H_0^2}\int^t_0 dt \frac{d}{dt}\dot{a}^3\;,
\end{equation}
where $V(0)$ is given through time dependent volume of region of interest dominated by dark energy, $V(t)=V(0) a^3=V(0) e^{3H_0t}$. Since in the semiclassical approximation the action is just the phase of the wavefunction, and for a topological term like the GB term the phase takes values in units of $2\pi$, we can write the total amount of the action contributed by the GB term at time $t$ , from Eq.~(\ref{GBFRW}), as 
\begin{eqnarray}
|\mathcal{S}_{GB}|=\frac{M_{Pl}^2 V(0) \dot{a}^3}{8\pi H_0^2}\;{\Bigg |}^t_0\equiv2\pi (N(t)-N(0))\;. 
\end{eqnarray}
Then, some sort of number density can be defined as 
\begin{eqnarray}
\label{edensity}
n\equiv\frac{N(t)}{V(t)}&=&\frac{ M_{Pl}^2} {16\pi^2 H_0^2}\left(\frac{\dot{a}}{a}\right)^3\nonumber\\
&=&\frac{H_0 M_{Pl}^2} {16\pi^2}\equiv \Lambda_{BZ}^3\;,\qquad \qquad \qquad \qquad 
\end{eqnarray}
which is time independent for the cosmological constant dominated space. Bjorken uses the term ``darkness'' for the quantity $N(t)$; we prefer to use the ``Gauss-Bonnet number''. 

Once we put in the numerical factors, the Bjorken-Zeldovich scale yields
\begin{flalign}
\label{BZscale}
\Lambda_{BZ}\sim 10\; \mbox{MeV} \quad \mbox{or} \quad l_{BZ}=\frac{1}{\Lambda_{BZ}}\sim 2\times10^{-12}\; \mbox{cm}\;.
\end{flalign}

$\Lambda_{BZ}$ appears to be the scale up to which the MM formalism is valid. Next, we will see how a length scale of the same size comes about as the compactification scale of six extra dimensions originated from TeV-scale gauge-gravity unification. Considering this as the picture above $\Lambda_{BZ}$, $N(t)$ can be interpreted as an effective quantity, revealed below $\Lambda_{BZ}$ upon integrated-over extra dimensions. This scenario, as we will see, accurately explains the factor $10^{120}$ in the MM action, as well. 
 
\section{Bjorken-Zeldovich scale from large extra dimensions}

In this section, we draw attention to an interesting incident regarding the onset of the scale of $10^{-12}$ cm from six compact extra (spatial) dimensions originated from TeV-scale gauge-gravity unification. If one imposes gauge-gravity unification at the TeV scale, the weakness of gravitational interactions can be explained via the existence of compact extra dimensions, large compared to the (inverse) TeV-scale~\cite{ArkaniHamed:1998rs,Antoniadis:1998ig,ArkaniHamed:1998nn,ArkaniHamed:2000eg,Antoniadis:1997zg,Accomando:1999sj,Antoniadis:1999bq}. 

For two test objects placed within a distance $r\gg R$, the gravitational potential is given as
\begin{equation}
V(r)\sim \frac{m_1 m_2}{M_U ^{n+2} R^n}\frac{1}{r}\;, \qquad (r\gg R)
\end{equation}
where $M_U$ is the unification scale of gauge and gravitational interactions, and $R$ is the compactification scale of the extra dimensions. Imposing the requirement to get the right (reduced) Planck mass through the identification $M_U ^{n+2} R^n=\overline{M}_{Pl}^2$, and assuming $M_U\sim 1$ TeV, we obtain
\begin{equation}\label{LEDscale}
R=\frac{l_U^{1+2/n}}{l_{Pl}^{2/n}}\sim2.0\times\left(2.4\right)^{2/n}\times 10^{30/n-17} \mbox{ cm}\;,
\end{equation}
where $l_U$ and $l_P$ are corresponding length scales for the TeV-scale and (reduced) Planck masses, respectively. As can be seen in Eq.~(\ref{LEDscale}), for n=6, we have
\begin{equation}\label{LEDlength}
R \sim 2.7\times10^{-12} \mbox{ cm}\sim l_{BZ}\;,
\end{equation}
a remarkable agreement with the Bjorken-Zeldovich length scale, given in Eq.~(\ref{BZscale}), revealed in the MacDowell-Mansouri formalism, discussed previously. 

Existence of six extra spatial dimensions compactified on a scale of $10^{-12}$ cm in the MM framework, as also noted in Ref.~\cite{ANDP:ANDP201300724}, could also explain the factor $\frac{M_{Pl}^2}{64\pi H_0^2}\;=\;10^{120}$, multiplying both the MM action given in Eq.~(\ref{action1}) and the GB term in the action given in Eq.~(\ref{action2}). Extending the internal symmetry of the general MM action from $SO(4,1)$ to $SO(10,1)$, and breaking the symmetry down to $SO(9,1)$, in analogy with Eq.~(\ref{action1}), the action symbolically becomes

%
\begin{flalign}
\mathcal{S}_{MM} \rightarrow \int d^4x \int  \sqrt{-\tilde{g}}\;dy_1...dy_6  \; (F)^5_{(i\;\overline{\mu})} \cdot \epsilon^{(i)}\cdot \epsilon ^{(\overline{\mu})}\;,
\end{flalign}  
where we suppress the complete version of the tensors, and the indices run over ten values instead of original four. We expect $\braket{F}$ to take a value around the order of the (square of the) energy scale that sets the strength for the effective 4 dimensional gravity, \textit{i.e.} the (reduced) Planck mass square, $\braket{F}\sim \overline{M}_{Pl}^2$ (or $\braket{F}\sim \overline{M}_{Pl}^2 \sim M_U^{8} R^6$ in the context of large extra dimensions picture discussed above). On the other hand, each integrated-over extra dimension contributes a factor in the order of the corresponding length scale, \textit{i.e.} $\int dy\sim l_{BZ}$. Therefore,
%
\begin{flalign}
\mathcal{S}_{MM} \;\;\;\rightarrow\quad &\underbrace{(\overline{M}_{Pl}^2 l_{BZ}^2)^3}&\hspace{-1cm}\int d^4x\; \sqrt{-g}\;\; (F)^2_{(a\;\mu)} \cdot \epsilon^{(a)}\cdot \epsilon ^{(\mu)}\;,\nonumber\\
& \sim10^{120}&
\end{flalign}
which accurately accounts for the factor $10^{120}$ in the MM action given in Eq.~(\ref{action1}).

In the case of the MM formalism with extra dimensions with a compactification size of $l_{BZ}\sim10^{-12}$ cm, $N(t)$ is an effective quantity arisen only when the extra dimensions are integrated over. Therefore, at distances smaller than $l_{BZ}$, or at energies above $\Lambda_{BZ}$, the expression for $n(t)$, given Eq.~(\ref{edensity}), which yields the BZ scale, is not well-defined. In other words, in this picture the MM description in 4 dimensions is valid up to $\Lambda_{BZ}$, and beyond that we have the 10 dimensional picture. Since the GB term is topological in 4D, there is no deviation from the usual Einstein-Hilbert gravity below $\Lambda_{BZ}$. On the other hand, above  $\Lambda_{BZ}$, the only modification is in the effective gravitational interaction (at distances smaller than $10^{-12}$ cm), since in this scenario only graviton and possibly other gravitational degrees of freedoms experience the extra dimensions but not the known (SM) fields.

\begin{figure}[t]
\begin{center}
\includegraphics[width=9cm]{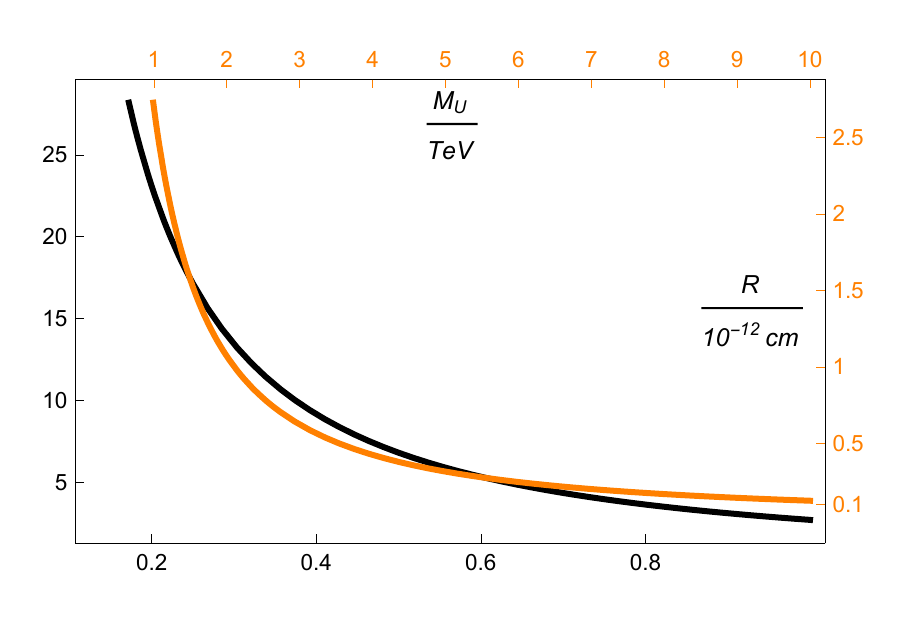}
\caption{\label{plot}
Compactification radius $R$ vs. quantum gravity scale $M_U$ in the case of six large extra dimensions. The black and orange plots denote the cases in which $M_U$ takes values below and above 1 TeV, respectively.}
\end{center}
\end{figure}

There are two main ways that the extra dimensions, in the context of the ADD model, would appear at the Large Hadron Collider (LHC). The first one is through the direct production of the graviton Kaluza-Klein (KK)-tower states, where the signal would appear as missing energy. The other signature would manifest itself via the exchange of virtual KK gravitons between the SM particles, which would give rise to enhancement in certain cross sections above the SM values~\cite{Csaki:2004ay,Rizzo:2010zf}. Currently at the LHC, the lower limit for the gauge-gravity scale in the ADD model with six extra dimensions is set as $M_U>2.6$ TeV with 95\% CL by the CMS experiment~\cite{Roy:2018tkg,Ghosh:2018cfr,CMS:2016fnh}, which translates into an upper bound on the corresponding length scale as $R<0.8\times 10^{-12}$ cm. This is still in the vicinity of the Bjorken-Zeldovich scale within an order of magnitude. Then, one wonders how robust the numerical agreement, given in Eq.~(\ref{LEDlength}), is against the value of $M_U$. As can be seen in Eq.~(\ref{LEDscale}), the outcome has some sensitivity against the value of $M_U$. Nevertheless, with a value of $M_U$ up to around 10 TeV,  we still get required length scale up to an order of magnitude, which is generally acceptable when a scale is under discussion, as displayed in Figure \ref{plot}. If, for instance, we take $M_U=5$ TeV, the corresponding length scale becomes $R=0.3\times 10^{-12} \mbox{cm}$; or, for $M_U= 10$ TeV, we have $R=0.1\times 10^{-12} \mbox{cm}\cong d_{\mbox{\scriptsize proton}}$.

\section{A~``see-saw" relation for the small cosmological constant}

In the effort to understand the smallness of the cosmological constant, several numerical relations among the energy scales have been noticed (or proposed) in the literature that mimic a see-saw-type suppression mechanism~\cite{Banks:1995uh,Cohen:1998zx,Kim:1999dc,Kiritsis:1999tx,ArkaniHamed:2000tc,Banks:2000fe,Chang:2001bm,Chang:2001kn,Barr:2001vh,Berglund:2002kr,Hsu:2004jt,Urban:2009wb,Urban:2009vy,Urban:2009yg,Chang:2010ir,Chang:2011jj}. 

In the case of the existence of an energy scale of $\sim10$ MeV, 
the relevant combination we notice is
\begin{equation}
\rho_{\Lambda}^{1/4}\overset{?}{=}M_{BZ}^3/M_{EW}^2\sim 10^{-3} \mbox{ eV}\;,
\end{equation}
where $M_{EW}\sim 1$ TeV.
It is not straightforward to devise a realistic model yielding such a relation, since this requires a contribution in the amount of $M_{BZ}^{12}/M_{EW}^8$ in Lagrangian. Nevertheless, this type of terms in the context of vacuum energy density contributions may be obtained in models where the cosmological constant problem is addressed by entertaining the possibility that the universe may be stuck in a false vacuum, split from the vanishing global vacuum in the amount of the cosmological constant~\cite{Garretson:1993kg,Barr:2001vh}. 

For instance, consider N real scalar fields $\phi_i$ which transform under the discrete group $\mathcal{G}_D= S_N\otimes \overline{\mathbb{Z}}$. $S_N$ is the permutation group, while $\overline{\mathbb{Z}}$ applies the operation ($\phi_i \rightarrow -\phi_i$, $\phi_j \rightarrow -\phi_j$, $\phi_n \rightarrow \phi_n$)  where $i\neq j$, $n\neq i, j$, and $i=1,...,N$. The renormalizable potential consistent with these symmetries,
\begin{flalign}
\mathcal{V}_{0}(\phi_i)=-m^2 \phi_i \phi_i +\lambda_1 \left(\phi_i^2\right)^2+\lambda_2 \left(\phi_i^2-\phi_j^2\right)^2_{i>j}\;,\nonumber\\
\end{flalign}
in which $m^2,\lambda_1,\lambda_2 > 0$, automatically exhibits an obvious ``accidental" symmetry under ($\phi_i \rightarrow -\phi_i$, $\phi_n \rightarrow \phi_n$), where $i\neq n$~\cite{Garretson:1993kg}. The latter symmetry, as we will see below, can be broken by higher order operators while the $\mathcal{G}_D$ symmetry is kept intact. The system has two sets of global vacua associated with the vacuum expectations values $\braket{\phi_i}=\pm m/\sqrt{2N\lambda_1}\equiv \pm v_{\phi}$. Note that in this kind of scenarios it is generally just \textit{assumed} that the system's global minimum is enforced to be vanishing by a yet-to-known mechanism, \textit{i.e.} $\mathcal{V}_{0}(\pm v_{\phi_i})=0$.  
In this toy example, each of the two sets of global minima consists of $2^{N-1}$ vacua, which are degenerate among themselves due to the exact $\mathcal{G}_D$ symmetry, while these two sets are degenerate to each other thanks to the accidental symmetry. The latter degeneracy between these two sets can be lifted by an operator added as a perturbation to the original potential. The potential with the first higher order operator that is consistent with the exact $\mathcal{G}_D$ symmetry yet breaks the accidental symmetry is    
\begin{eqnarray}
\mathcal{V} ^{eff}(\phi_i)=\mathcal{V}_{0}(\phi_i)+\frac{\lambda}{M_{EW}^{N-4}}\phi_1 \phi_2 ..\phi_N+\cdots\;.
\end{eqnarray}
where $(\cdots)$ denotes the rest of the terms up to the required order. Note that we maintain the idea from the previous section that $M_{EW}$ is the most fundamental scale of Nature. The example here could be trivially incorporated in the extra dimensional picture as long as the scalars here are assumed to be localized to the usual 4 dimensions. For a system with $N=12$ and $\braket{\phi_i}\sim M_{BZ}$, we obtain the desired result that the false vacuum is split from the vanishing true vacuum in the amount of $M_{BZ}^{12}/M_{EW}^8$. Note that the other, symmetric, terms in the effective potential introduce small oscillations around the either minimum and they are unlikely to be large enough to leap the system over the barrier separating the false and true vacua. Therefore, assuming the universe is in the false vacuum with $E=\rho_{\Lambda}^{1/4}$, then the only way for a transition to the vanishing true vacuum is barrier penetration by quantum tunnelling. 

At this point then, the next issue to address is the stability of the system in the false vacuum~\cite{Coleman:1977py,Callan:1977pt,Coleman:1980aw}. In order for a bubble of the true vacuum within the the false vacuum to be energetically favourable to expand after nucleation, instead of shrinking away, it is required that $dE_{bubble}/dR<0$. The energy of the bubble is given as 
\begin{equation}
E_{bubble}=-\frac{4}{3}\pi R^3 \rho_{\Lambda}+4\pi R^2 \sigma\;,  
\end{equation}
where $\sigma\sim v_{\phi}^3\sim M_{BZ}^3$ is the surface tension. The critical radius can be found via $dE_{bubble}/dR=0$ as 
\begin{equation}
R_{critical}\sim 10^{28} \mbox{ cm} \sim H_0\;,
\end{equation}
which means that for a bubble in a cosmological constant dominated false vacuum to nucleate and grow, it must be formed with the size of the observable universe, or larger. Therefore, the decay of the vacuum, and hence the stability, in such a scenario is not an issue for concern, which is generally the case in similar scenarios~\cite{Garretson:1993kg}. 

\section{Small neutrino mass via a see-saw type mechanism at the BZ scale}
When a new scale is under discussion, one of the questions that comes to mind is the possibility of a (some sort of) see-saw mechanism which utilizes a relevant combination of the scales in the theory to explain the smallness of the neutrino mass. The relation $M_{BZ}^3/M_{EW}^2\sim 10^{-3}$ eV is intriguing from this point of view as well, since it is in the vicinity of (at least one of the) the neutrino masses. Next, we will work on a hypothetical scenario just as an illustration of obtaining this combination in a model.    

Consider a hidden sector with a QCD-like gauge interaction, where the symmetry group is $\mathcal{G}_h \equiv SU(N)$. Besides the corresponding gauge bosons, consider a real scalar field $\phi$ and a Dirac fermion $F$ (for each family), both of which transform in some representation of $\mathcal{G}_h$, where the fermion does so vectorially (non-chirally). We assume that $F$ has a confining scale of order $10$ MeV and the SM fields are not charged under $\mathcal{G}_h$. The SM connects to the hidden sector through a portal coupling between the Higgs and the scalar $\phi$. We also assume a discrete $\tilde{Z}_2$ symmetry that transforms $F$, $\phi$, and the neutrinos in the following way. $F_{L(R)}\rightarrow \pm F_{L(R)}$, $\nu_{L(R)}\rightarrow \pm \nu_{L(R)}$, and $\phi \rightarrow - \phi$. Therefore, there are no mass or Yukawa-type terms (via the Standard Model Higgs) allowed at the tree level. The Yukawa-type terms involving the scalar $\phi$ do not contribute as mass terms at tree-level either, since we assume that $\braket{\phi}=0$ so that the $SU(N)$ symmetry remains unbroken.  However, a mass term can be induced through the effective operator 
\begin{equation}  
\mathcal{L} ^{eff} \supset 
\frac{c_{\nu}}{\Lambda^2} \overline{\nu}\nu \overline{F}{F}\;,
\end{equation}
 which is induced by integrating out the scalar field. We assume that a condensate forms,  due to the possible nonperturbative characteristic of the $SU(N)$ vacuum, at $\sim M_{BZ}$, breaking the $\tilde{Z}_2$ symmetry; $\braket{\overline{F}{F}}\sim f^3 \sim (a\cdot M_{BZ})^3$, $\Lambda\sim m_{\phi}\sim \Lambda_{EW} \sim 1$ TeV.  The scalar $\phi$ gets its mass via the portal coupling to the SM Higgs, \textit{i.e.} $\lambda \phi^2 H^{\dag}H$, which justifies its electroweak-scale mass. The extra fermion F acquires its mass via coupling to the condensate through the corresponding dimension-6 operator that yields a mass value on the order of the neutrino mass. The effective neutrino mass\footnote{We also note that in the context of large extra dimensions as well, the small neutrino mass can be obtained with a see-saw-like mechanism provided that the SM singlet right handed neutrino $v_R$, unlike the SM fields including the left handed neutrino $v_L$, is not confined in the 3-brane but instead lives in the bulk experiencing the extra dimensions~\cite{Dienes:1998sb,ArkaniHamed:1998vp}. The suppression of the Dirac mass $m_D$, in this picture, is given as in
 \begin{equation}
 m_{\nu}\sim \left(\frac{N}{2\pi}\right)^{n/2} \frac{m_D}{\left(M_U R\right)^{n/2}}\;, 
 \end{equation}
 for a general $\mathbf{ \mathbb{Z}}_N$ orbifold on which $n$ extra dimensions are compactified. For instance, for $n=6$ on a $\mathbf{ \mathbb{Z}}_2$ orbifold, with the quantum gravity scale of $M_U\sim 1$ TeV and hence $R=l_{BZ}\sim10^{-12}$ cm, the physical neutrino mass is obtained as
 \begin{equation}
 m_{\nu}\sim m_D\times 10^{-16}\;.
 \end{equation}
For $m_D\sim1$ TeV, for example, $m_{\nu}\sim10^{-4}$ eV.} in this scenario becomes  
 \begin{equation}
 \label{neutrinomass}
 m_{\nu}=(c_{\nu}a^3)\; M_{BZ}^3/M_{EW}^2\lesssim 10^{-2}\mbox{ eV}\;,
 \end{equation}
 provided that $c_{\nu}a^3\lesssim 10$.\footnote{The idea that condensates may give masses to particles should be treated with caution in case of existence of confinement. For instance, it is argued in Refs.~\cite{Brodsky:2008xm,Brodsky:2008be,Brodsky:2010xf,Brodsky:2012ku} that the QCD condensates have spatial support within hadrons and do not extend throughout the whole space. Note also that this is why it may not be appropriate to count the QCD vacuum condensates as contributions to the effective cosmological constant, as pointed out also in Refs.~\cite{Brodsky:2008xu,Brodsky:2009zd}. }


\section{More on the relevance in cosmology}

A scale around $10$ MeV might be relevant also in terms of the thermal history of the universe. It is an energy scale considerably close to $T\sim 1-2$ MeV below which the weak interactions freeze out; the reaction rate $\Gamma\sim G_F^2 T^5$ drops below the expansion rate $H\sim \sqrt{g^*} T^2 /M_{Pl}$, where $g^*$ is  given as $g^*=g_b+(7/8)g_f$ and $g_b\;(g_f)$ denotes the total number of the effective bosonic (fermionic) degrees of freedom at around the background temperature $T$. Consequently, primordial neutrinos and possibly cold dark matter -if it exists- decouple from the rest of the matter, and  the ratio of neutrons to protons freezes out. Any increase from the known radiation density would bring forward the Big Bang Nucleosynthesis (BBN) and hence would cause a larger Helium abundance in the present universe~\cite{Carroll:2004st,Weinberg:2008zzc,Langacker:2010zza}. Therefore, if there is some unrevealed physics associated with such a scale of $10$ MeV, there may have direct implications on our understanding of BBN, which is consistent with the current observations on the primordial abundances of light elements. 

Since the effective MM action in 4D reveals the Einstein gravity with a cosmological constant and the Gauss-Bonnet term that does not have any effects in the equations of motion in 4D, the formalism at first sight only defines the graviton. This seemingly does not cause any problem in terms BBN since the gravitational interaction rate, as well known, is significantly suppressed compared to the expansion rate, \textit{i.e.} $\Gamma \sim G_P^2 T^5 \ll H$, at $T\sim 1$ MeV. However, this is the case only if there is no any other relevant degrees of freedom obtained from the original action based on $SO(4,1)$, in addition to the terms given in Eq.~(\ref{action2}). Recall that the generalized connection $A^{AB}_{\mu}$ living in $SO(4,1)$ has 40 components. As also mentioned in Ref.~\cite{ANDP:ANDP201300724}, one may wonder whether some of these degrees of freedom can be identified with the (bosonic) degrees of freedom of the SM\footnote{See, for instance, Refs.~\cite{Lisi:2007gv,Nesti:2009kk,Lisi:2010td,Neeman:1979wp,Fairlie:1979at,Fairlie:1979zy,Neeman:1990tfh,Neeman:2005dbs,Chamseddine:2010ud,Chamseddine:2012sw,Chamseddine:2013rta,Aydemir:2013zua,Aydemir:2014ama} for various geometric approaches to the Standard Model and beyond.}. Then, several leftover terms may possibly define additional light degrees of freedom. One may expect at first that the relevant interactions are supposed to be suppressed, similar to the case with gravitons. However, one should not forget the enormous factor of $10^{120}$ multiplying the action in 4D, possibly arisen due to the integrated over extra dimensions. If such identifications related to the SM are possible, then it is probably because of this large factor, and the same factor may amplify some interactions regarding these new light degrees of freedom, making them interact frequently enough to be in equilibrium at $T\sim 1$ MeV. Then, the model would be in tension with the constraints coming from BBN. 

A comment is in order on the GB number density in the MM framework, $n(t)$, earlier (than $1/\Lambda_{BZ}$) in the thermal history when the universe is dominated by radiation. Bjorken, in Ref.~\cite{ANDP:ANDP201300724}, interprets the energy where the GB number density is Planckian, \textit{i.e.} $n(t)\sim M_{Pl}^3 $, as the cut-off for the MM description above which some modification is necessary. By using the time dependent expression, given in Eq.~(\ref{edensity}), and the equations 
 \begin{equation}
 H^2=8\pi\rho/(3 M_{Pl}^2)\quad \mbox{and}\quad \rho_R(t)=\pi^2 g^* T^4/30\;,
 \end{equation}
which follow from the first FRW equation and the Stefan-Boltzmann law for species in thermal equilibrium, respectively,  
it is found that this critical energy turns out to be $T_c\cong60\mbox{ MeV}\sim 6\Lambda_{BZ}$ when only the known-relevant relativistic degrees of freedom are active, \textit{i.e.} $g^* =43/4$. However, we note that this interpretation may only be valid if we do not admit extra dimensions in the framework, as we discussed previously. This is because at distances smaller than the BZ length $10^{-12}$ cm, or at energies above BZ energy $10$ MeV, the expression for $n(t)$, given Eq.~(\ref{edensity}) is not well-defined. Note also that although above $10$ MeV additional gravitational degrees of freedom may arise due to the large internal group, $SO(10,1)$, of the theory with 6 extra dimensions, the corresponding interactions are expected to be suppressed, as in the case with gravitons, and not to have noticeable effects on the energy density. Recall that the factor $10^{120}$, which may amplify the couplings, appears only when the theory is integrated over the extra dimensions.


\section{Summary and Discussions} 

In this paper, we aim to bring into attention the possibility of a gravitational scale at around $\sim 10$ MeV, or $10^{-12}$ cm, induced in the MacDowell-Mansouri (MM) extension to the Einstein-Cartan formalism in 4D, due to the topological Gauss-Bonnet term in the action, as suggested by Bjorken~\cite{ANDP:ANDP201300724}.

\textit{First}; we point out that a scale of the same size, $\sim10^{-12}$ cm, naturally comes about in the context of large extra dimensions, originated from TeV-scale quantum gravity, as the compactification scale, if the number of extra spatial dimensions is set to six.\footnote{Recall that existence of six compact extra dimensions is quite familiar from the string theory perspective~\cite{Schwarz:1974ix,Scherk:1975fm,Candelas:1985en}.}~Apparently, these two approaches can be combined, where the 4-dimensional MM formalism is the effective theory after the six extra dimensions are integrated over, which also explains the factor $10^{120}$ in the MM action. \textit{Second}; we discuss that existence of such a scale may play a role in the smallness of the cosmological constant and the neutrino mass; to this end, we refer to some toy models as illustrations that generate a seesaw-type suppression mechanism. Note that we do not claim in this paper that this mechanism can directly be accommodated in the MM formalism. Rather, the main intention of this paper should be taken as an attempt to point out several coincidences regarding a scale at around $10$ MeV, or $10^{-12}$ cm; the possibility that they are non-accidental deserves attention. Finally, we comment on possible implications in cosmology in the context of Big Bang Nucleosynthesis.     
 \raggedbottom     
 
We note that if Nature contains six extra dimensions with a compactification scale of $10^{-12}$ cm, it raises the question why no Kaluza-Klein excitations with masses with a starting value of $\sim10$ MeV have been observed so far. However, their elusiveness would not be unanticipated since these modes are expected to be relatively suppressed~\cite{Han:1998sg}.  

Currently at the LHC, the ADD model with six extra dimensions is excluded with 95\% CL for values $M_U\leqslant2.6$ TeV by the CMS experiment~\cite{Roy:2018tkg,Ghosh:2018cfr,CMS:2016fnh}, equivalent to an upper bound on the corresponding compactification length scale, $R<0.8\times 10^{-12}$ cm, which is in the ballpark of the Bjorken-Zeldovich scale within an order of magnitude. By the time the LHC searches are completed, we will have a compelling answer on the TeV-scale ADD model and hence on the MM-LED picture discussed in this paper. Note that a negative result along these lines does not necessarily invalidate Bjorken's original proposal in Ref.~\cite{ANDP:ANDP201300724}, which is not obliged to connect to the TeV-scale ADD model, and yet which can still include six compact extra dimensions. The most definitive answer regarding Bjorken's proposal will come from the gravitational inverse-square-law experiments. The current upper bound for the size of such extra dimensions, through the deviation in the effective gravitational interaction, is $0.01$ cm~\cite{Kapner:2006si}.
 

\section*{Acknowledgements} 

This work is supported in parts by the National Natural Science Foundation of China (NSFC) under Grant No.~11505067 and the Swedish Research Council under contract 621-2011-5107. We would like to thank James D. Bjorken and Djordje Minic for their comments and suggestions regarding the manuscript. \vspace{0.5cm}

\raggedright
\bibliography{BZ-scale}{}
\bibliographystyle{apsrev4-1}

\end{document}